\documentclass[aps,prb,10pt,twocolumn,superscriptaddress,floatfix,amsmath,amssymb]{revtex4-2}
\usepackage{silence}
\usepackage{amsmath}
\usepackage{amssymb}
\usepackage{graphicx}
\usepackage{float}
\begin{document}
\title{Strain-dependent one-dimensional confinement channels in twisted bilayer 1T$'$-WTe$_2$}

\author{S. J. Magorrian}
\affiliation{Department of Physics, University of Warwick, Coventry, CV4 7AL, United Kingdom}

\author{N. D. M. Hine}
\email{n.d.m.hine@warwick.ac.uk}
\affiliation{Department of Physics, University of Warwick, Coventry, CV4 7AL, United Kingdom}

\begin{abstract}
   The low symmetry and anistropic lattice of 1T$'$ WTe$_2$ is responsible for the existence of parallel one-dimensional channels in the moiré patterns of twisted bilayers. 
   This gives the opportunity to explore moiré physics of a different nature to that widely observed in twisted bilayers of materials with hexagonal symmetries.
   Here, we combine plane-wave and linear-scaling density functional theory calculations to describe the electronic properties of twisted bilayer 1T$'$  WTe$_2$. 
   For a small change in the lattice parameters of the constituent 1T$'$ WTe$_2$ monolayers, we find a substantial moiré-induced striped electrostatic potential landscape in the twisted bilayer, with a peak-to-trough magnitude $>$200~meV.
\end{abstract}

\maketitle

\section{Introduction}
The novel and interesting physics arising from the long-range periodicity present in the moiré patterns of twisted bilayers of two-dimensional (2D) materials with hexagonal symmetry, such as graphene\cite{Cao2018, Yankowitz2019, Sharpe2019, Serlin_2020} and the transition metal dichalcogenides (TMDCs)\cite{Lin_2018, Wang2020, Tang_2020, Regan2020, Quan_2021}, has given rise to extensive research into the physical phenomena hosted in these systems.
Typical twisted bilayers of TMDCs comprised of monolayers exfoliated from 2H-polytype crystals feature moiré superlattices with hexagonal patterns and symmetries inherited from the symmetries of their constituent monolayers\cite{Ferreira2021_APL}.

\begin{figure}
    \centering
    \includegraphics[width=1.00\linewidth]{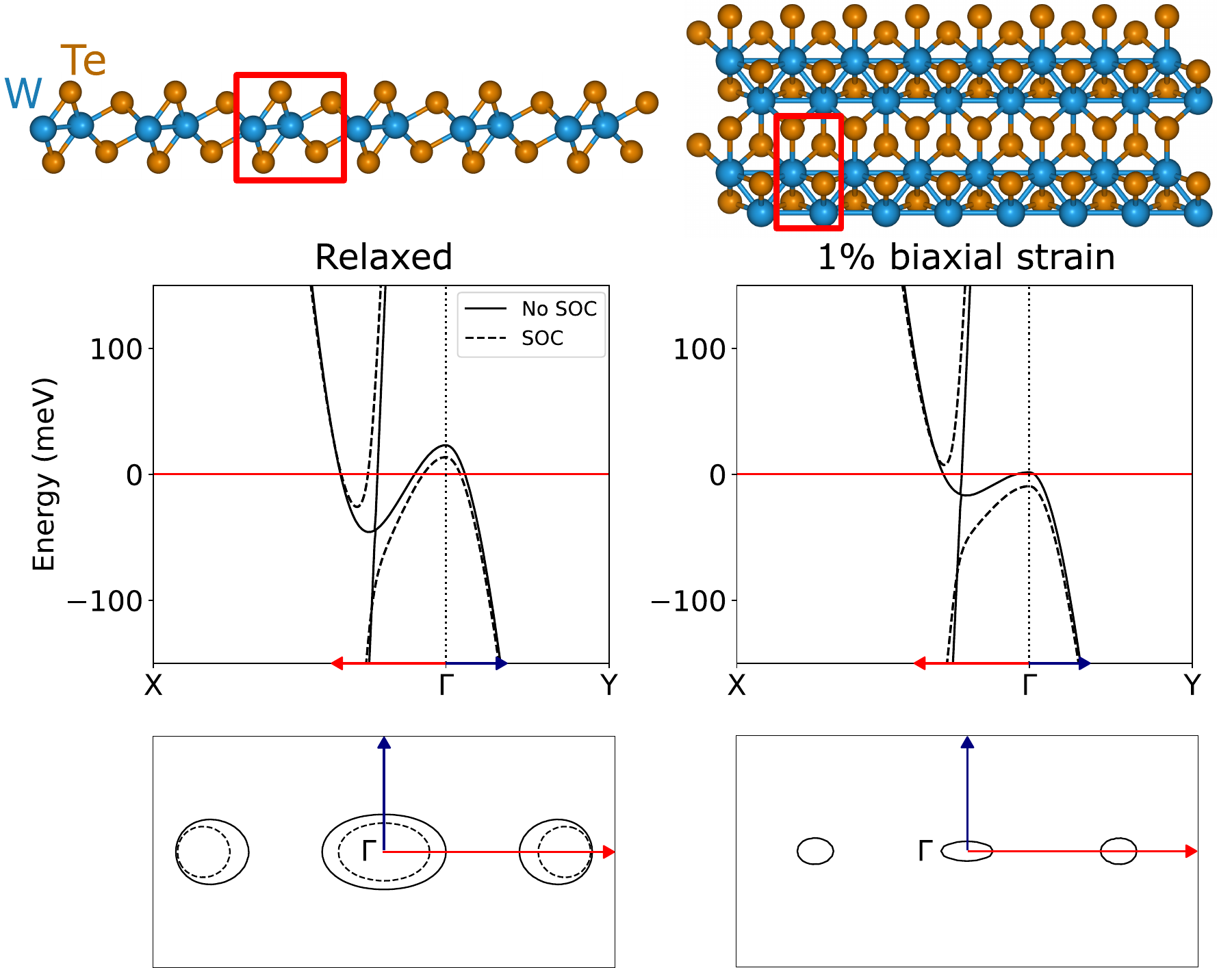}
    \caption{Upper panels: side and top views of monolayer 1T$'$-WTe$_2$, with red boxes highlighting the primitive unit cell of the system. Middle panels - band structures, of relaxed and 1\% biaxially strained monolayer 1T$'$ WTe$_2$. Solid and dashed lines are for calculations neglecting and including spin-orbit coupling (SOC), respectively. Lower panels - Fermi surfaces extracted from the band structures shown.}
    \label{fig:monolayer}
\end{figure}

\begin{figure}
    \centering
    \includegraphics[width=1.00\linewidth]{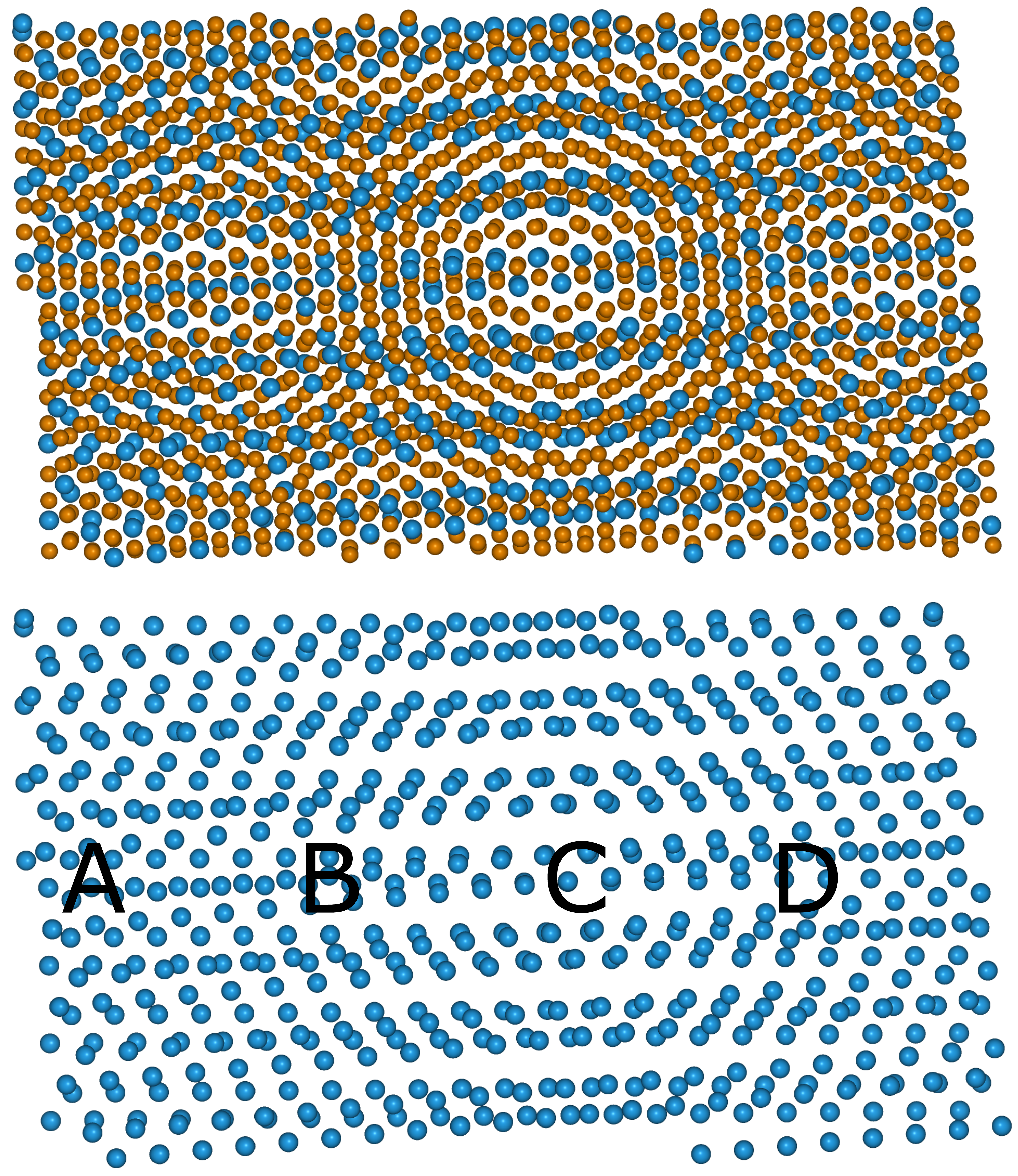}
    \caption{Upper panel: supercell twisted bilayer 1T$'$-WTe$_2$ with twist angle $\theta=4.66^{\circ}$ used in LS-DFT calculations. Lower panel: structure of twisted bilayer showing only tungsten atoms, highlighting one-dimensional channel in the moiré pattern. A-D mark regions with local atomic registries for which aligned bilayer band structures are presented in Fig. \ref{fig:aligned_bands}.}
    \label{fig:structure}
\end{figure}

More recently, attention has turned to those 2D materials which have different, usually lower, symmetries. These include the GeSe\cite{Kennes2020}, phosphorene\cite{Soltero2022}, and the 1T$'$ phase of the TMDCs. Among the TMDCs (MX$_2$, M=Mo,W; X=S,Se,Te) the hexagonal 2H polytype is generally energetically favoured over 1T$'$\cite{Duerloo2014}, with only WTe$_2$ having the 1T$'$ phase as the most stable.
This phase of WTe$_2$, the structure of which is shown in the top panels of Fig. \ref{fig:monolayer}, features distorted atomic layers, and a rectangular primitive cell. 
Owing to the reduced symmetry and anistropy of the distorted monolayer, twisted bilayer 1T$'$ WTe$_2$ has a qualitatively different moiré pattern to that familiar in twisted bilayers of the 2H phase.
This pattern is demonstrated in the moiré supercell of twisted bilayer 1T$'$ WTe$_2$ with twist angle $\theta=4.66^{\circ}$, illustrated in Fig.~\ref{fig:structure}.
The moiré supercell is rectangular, inherited from the rectangular primitive cell of monolayer 1T$'$ WTe$_2$, while the chains of the tungsten sublattices form one-dimensional stripes in the moiré pattern. 
Transport experiments on twisted bilayer 1T$'$ WTe$_2$ with a twist angle $\sim 5^{\circ}$\cite{Wu_WTe2} revealed highly-anistropic transport properties, with conductance across the stripes in the moiré pattern exhibiting a power-law scaling consistent with the formation of a 2D array of 1D electronic channels hosting Luttinger liquid behavior.

The monolayer band structure of 1T$'$ WTe$_2$, which we summarize via DFT in Fig.~\ref{fig:monolayer}, features a hole pocket at $\Gamma$, flanked by two electron pockets. 
In the absence of spin-orbit coupling (SOC), there is a protected band crossing near the minimum of the electron pocket.
SOC lifts this degeneracy, depending on the overlap between the electron and hole pockets this results in either a semimetallic or gapped band structure. 
Due to an inversion between the two bands involved 1T$'$ WTe$_2$ has long been predicted theoretically to host in the monolayer gapped quantum spin Hall (QSH) insulating behavior\cite{Qian2014, Zheng2016}. 
Following the theoretical predictions of QSH behavior in 1T$'$ WTe$_2$, experimental confirmation of inverted bands has been obtained, with a spin-orbit induced gap and a conducting edge state\cite{Fei2017, Tang2017,PhysRevB.96.041108, Cucchi2018, Wu2018, Shi2019} observed, while strain tunability between a semimetallic phase and a gapped phase has been reported\cite{PhysRevLett.125.046801}.
Evidence has also been found\cite{Jia_2021,Sun_2021} indicating that the gap-opening mechanism in monolayer 1T$'$ WTe$_2$ is the formation of electron-hole pairs, forming an excitonic insulator at charge neutrality, as opposed to the system being a simple band insulator or electron localization being responsible for the formation of a gap.

Few-layer films and bulk crystals of 1T$'$ WTe$_2$ have been observed to exhibit ferroelectric polarization despite being metallic\cite{Fei2018,Sharma2019}, with switching achieved via a small relative in-plane shift between the layers\cite{Yang2018,Liu2019}.
In the bulk the topologically-rich nature of 1T$'$ WTe$_2$ appears in its character as a type-II Weyl semimetal\cite{Lin_TIIWeyl_2017}.


In this work, we carry out first principles modelling of twisted bilayers of 1T$'$ to investigate the properties of the low-symmetry striped moiré patterns of twisted bilayer 1T$'$ WTe$_2$.
These simulations combine plane-wave density functional theory (DFT) calculations for primitive-cell aligned bilayers of WTe$_2$ with a direct linear-scaling DFT calculation for a large low-strain supercell of twisted bilayer WTe$_2$ (illustrated in Fig. \ref{fig:structure}), with a twist angle of 4.66$^{\circ}$.
Comparison of the electrostatic potential of the twisted bilayer with those of isolated monolayers reveals the twisted bilayer reveals an instability to a small adjustment in the lattice of the constituent monolayers. 
On application of 1\% biaxial strain a confining potential wave on the scale of the one-dimensional striped moiré pattern forms, with a peak-to-trough magnitude $>$200~meV.
We calculate spectral functions for the twisted bilayer unfolded and projected onto the primitive cells of the constituent monolayers, revealing a substantial disruption to the band structure due to the confining potential wave in the strained case.
Finally, we calculate using plane-wave DFT band structures for aligned primitive bilayers of 1T$'$ WTe$_2$ with stackings corresponding to the local atomic registries found in different regions of the moiré supercell of the twisted bilayer.
These show notable changes in topology both on changing the local bilayer stacking order, and on application of a small strain.


\section{Methods}
To obtain a relaxed monolayer geometry for 1T$'$ WTe$_2$ we carried out plane-wave DFT calculations using the QUANTUM ESPRESSO package\cite{Giannozzi2009, Giannozzi2017}.
The generalized gradient approximation of Perdew, Burke, and Ernzerhof is used\cite{PBE} with projector-augmeneted wave (PAW) pseudopotentials generated via a fully-relativistic calculation\cite{DalCorso2014} with spin-orbit coupling taken into account.
For the plane-wave DFT calculations neglecting spin-orbit coupling (SOC) we use pseudopotentials from the JTH dataset dataset\cite{Jollet_2014}, which are generated using a scalar-relativistic approximation.
Cutoffs of 80~Ry and 720~Ry are used for the wavefunctions and the density, respectively.

Direct atomistic simulation of twisted bilayers of two-dimensional materials can be challenging, due to the large number of atoms required to form a moiré supercell without needing to apply unreasonably large strains to enforce commensurability for a periodic calculation.
For our large-scale simulations of twisted bilayer 1T$'$ WTe$_2$ we therefore employ linear scaling density functional theory (LS-DFT), as implemented in the ONETEP code\cite{ONETEP}.
For all ONETEP calculations we use a psinc-grid energy cutoff of 800~eV, a cutoff radius of 13~$a_0$ for the non-orthogonal generalised Wannier functions (NGWFs), with 13 NGWFs on W atoms and 4 on Te atoms.
No truncation is applied to the density kernel.
Repeated images of the slab are placed 60~\AA~apart, with a cutoff applied to the Coulomb interaction in the out-of-plane direction to suppress any interaction between the repeated slabs\cite{Rozzi2006,Hine2011}.
We use the projector-augmented wave (PAW) method\cite{Hine_2016} with potentials from the JTH dataset\cite{Jollet_2014}.
We calculate unfolded spectral functions\cite{Popescu2012} for the twisted bilayers following a methodology which adapts the approach to the NGWF representation, as described previously\cite{Constantinescu2015}. 

To form the crystal structure of the twisted bilayer, we take the relaxed monolayer geometry found using plane-wave DFT as described above, constructing a twisted bilayer using rigid copies of this monolayer, since at the twist angle we consider the moiré period is short enough that atomic reconstruction will be minimal\cite{Yuan2023}.
We look for a supercell with a twist angle close to that measured in 
the aforementioned transport measurements\cite{Wu_WTe2}.
We find a suitable supercell with a twist angle $\theta = 4.66^{\circ}$, close to the twist angles of the devices measured in Ref. \onlinecite{Wu_WTe2}.
The supercell contains 1848 atoms, and a maximum magnitude component of the strain tensor of 0.33\%.
 For accurate energetics of the interlayer interactions, we employ the optB88-vdW\cite{Klime_2009} non-local van der Waals functional.
 The interlayer distance $d$ is found by minimizing energy as a function of $d$ between rigid monolayers.
 The result in this system is $d=7.75$\AA, which is in line with expectations, being around 10\% greater than the layer separation in bulk WTe$_2$ (c=7.035~\AA \cite{Brown1966}).


For interpretation of the stacking-dependence of the band
structure, we also perform plane-wave DFT band structures for aligned primitive bilayers of 1T'-WTe$_2$ for a selection of in-plane displacements.
The same interlayer distance ($d=7.75$~\AA) from the twisted bilayer is carried over for these calculations, since at this relatively small moir\'e size, the interlayer distance is expected to remain roughly constant despite the changes in stacking through the unit cell.

\section{Results}
\subsection{Strain-dependent confining potential}

\begin{figure}
    \centering
    \includegraphics[width=\linewidth]{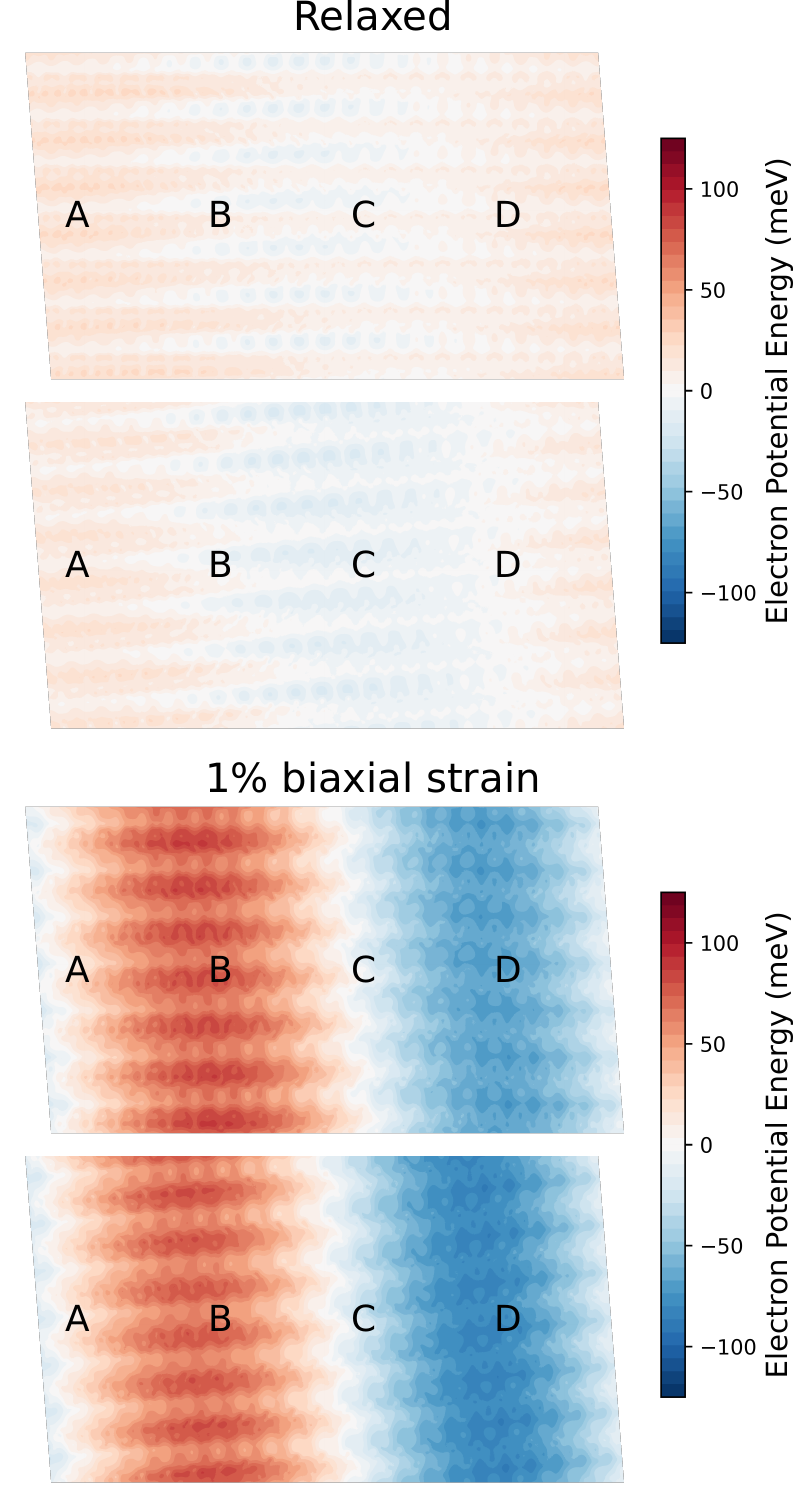}
    \caption{Upper two panels: Confining potentials in twisted bilayer 1T$'$-WTe$_2$, calculated as the difference between electron potential energy of the twisted bilayer and the sum of the potentials from isolated monolayers with atomic positions taken from the bilayer. Upper and lower panels show potentials at mean planes of the two layers. Supercell is that shown in Fig. \ref{fig:structure}. Letters ABCD indicate regions in which the local atomic registry approximates the aligned bilayer structures for which bands are calculated in Fig. \ref{fig:aligned_bands}. Lower two panels: same quantities, but for 1\% strain applied to constituent monolayers.}
    \label{fig:confining_potential}
\end{figure}

To extract from our LS-DFT calculations for the twisted bilayer the changes in potential felt by electrons due to the moiré superlattice, we first perform calculations for each of the two isolated monolayers individually, with their atomic positions as in the twisted bilayer, then perform a calculation of the twisted bilayer.
We then take the difference between the electrostatic potential in the twisted bilayer and the sum of the potentials of the isolated monolayers.
The resulting quantity is specifically the change in potential due to the interaction and hybridization between states on the two layers.
We show maps of the in-plane position-dependence of this potential for $z$ values corresponding to the mean planes of each layer in Fig.~\ref{fig:confining_potential}. 

In TMDC bilayers marginally twisted from 3R stacking\cite{Ferreira2021} it is known that equal-and-opposite potentials arise due to charge transfer between the layers, possible because of the layer-asymmetry of the bilayer.
Twisted bilayers of 1T$'$-WTe$_2$ behave rather differently: here the approximate local inversion symmetry of nearly-parallel bilayer stacking requires the sign of the potential to be the same on the two layers.
Fig.~\ref{fig:confining_potential} we see this confining potential for two cases: using constituent monolayers relaxed as described above, and a second case using monolayers biaxially strained by 1\%. 
For the unstrained case we see only small changes ($\sim$ 10s of meV) in the local electrostatic potential on formation of a twisted bilayer. 
On application of 1\% strain the regime changes drastically, with the emergence of a striped potential landscape with a peak-to-trough strength $\sim$200~meV.
This striped pattern runs parallel to the stripes in the moiré pattern formed by varying displacement of the axes of the chains of tungsten atoms.
The relative arrangements of the tellurium lattices, and local displacements along the tungsten chains, have minimal effect on the confining potential difference.

\subsection{Unfolded spectral functions}
\begin{figure}
    \centering
    \includegraphics[width=0.95\linewidth]{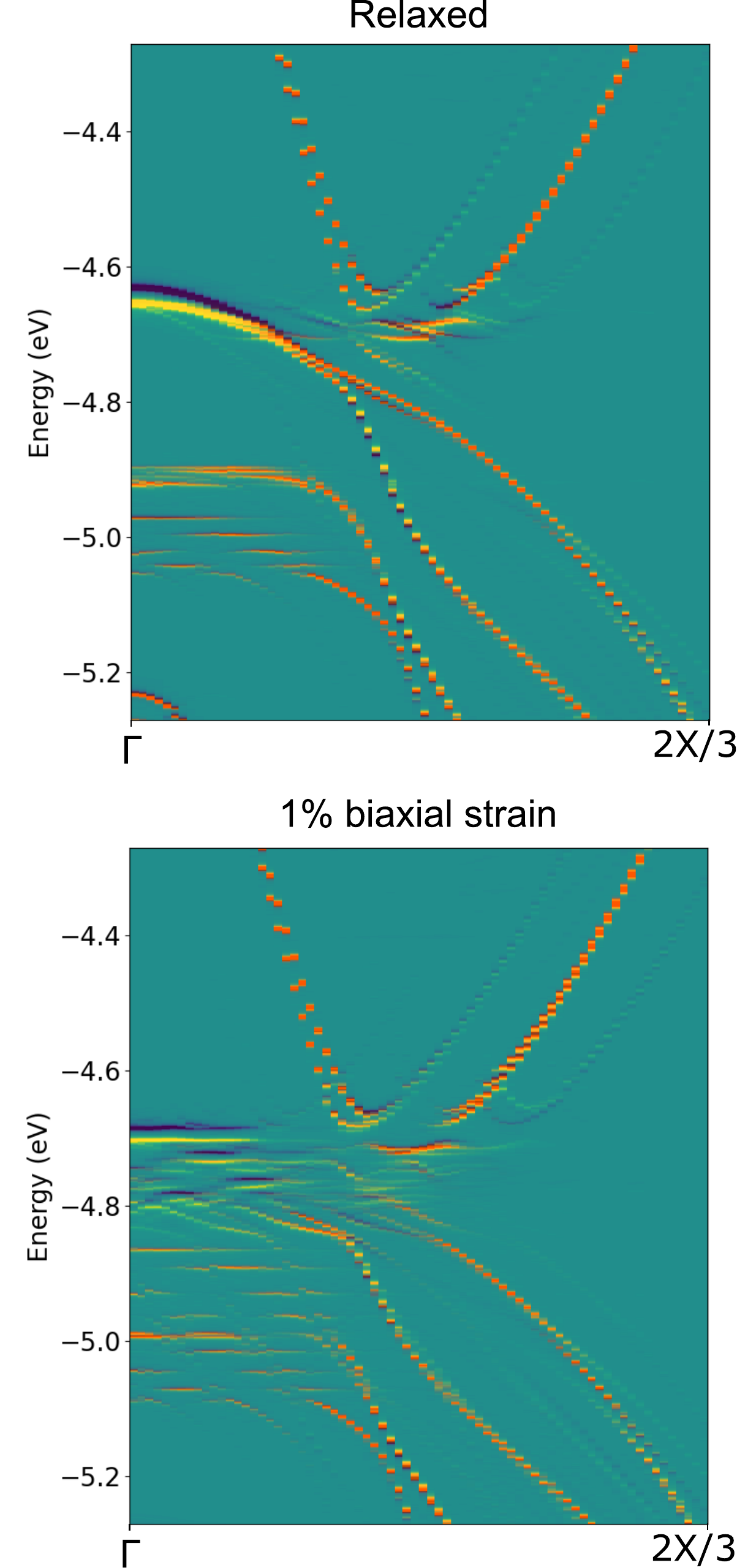}
    \caption{Spectral function of twisted bilayer 1T$'$ WTe$_2$, unfolded and projected onto primitive cells of each monolayer - yellow and blue bands correspond to projection on different layers, and add to orange where bands coincide. Spin-orbit coupling is taken into account perturbatively. Upper and lower panels are constructed from unstrained and 1\% biaxially strained monolayers, respectively.}
    \label{fig:specfuncs_unfolded}
\end{figure}
Further insight can be obtained by plotting the spectral function for the twisted bilayer, unfolded and projected onto the primitive cells each of the two layers, as shown in Fig. \ref{fig:specfuncs_unfolded}.
There are some differences between the unfolded bands on each layer, mainly due to the strain required that needs to be applied to one layer to form a commensurate unit cell.
These differences are, however, small compared to other effects, and are only apparent in certain regions of the band structure, for example near the local maximum at the $\Gamma$ point.
Substantial changes to the bilayer band structure are apparent. The minima of the electron pockets become detached from their bands, while below the valence band edge, the second band, which is particularly flat, is split into a standing wave-like structure.
This effect is even stronger for the 1\% biaxially strained case, where strong confining potential explored above causes the multiplicity of flat bands reaches to the top of the hole pockets.
\subsection{Strain induced Fermi-surface changes in aligned bilayers}
\begin{figure*}
    \centering
    \includegraphics[width=\linewidth]{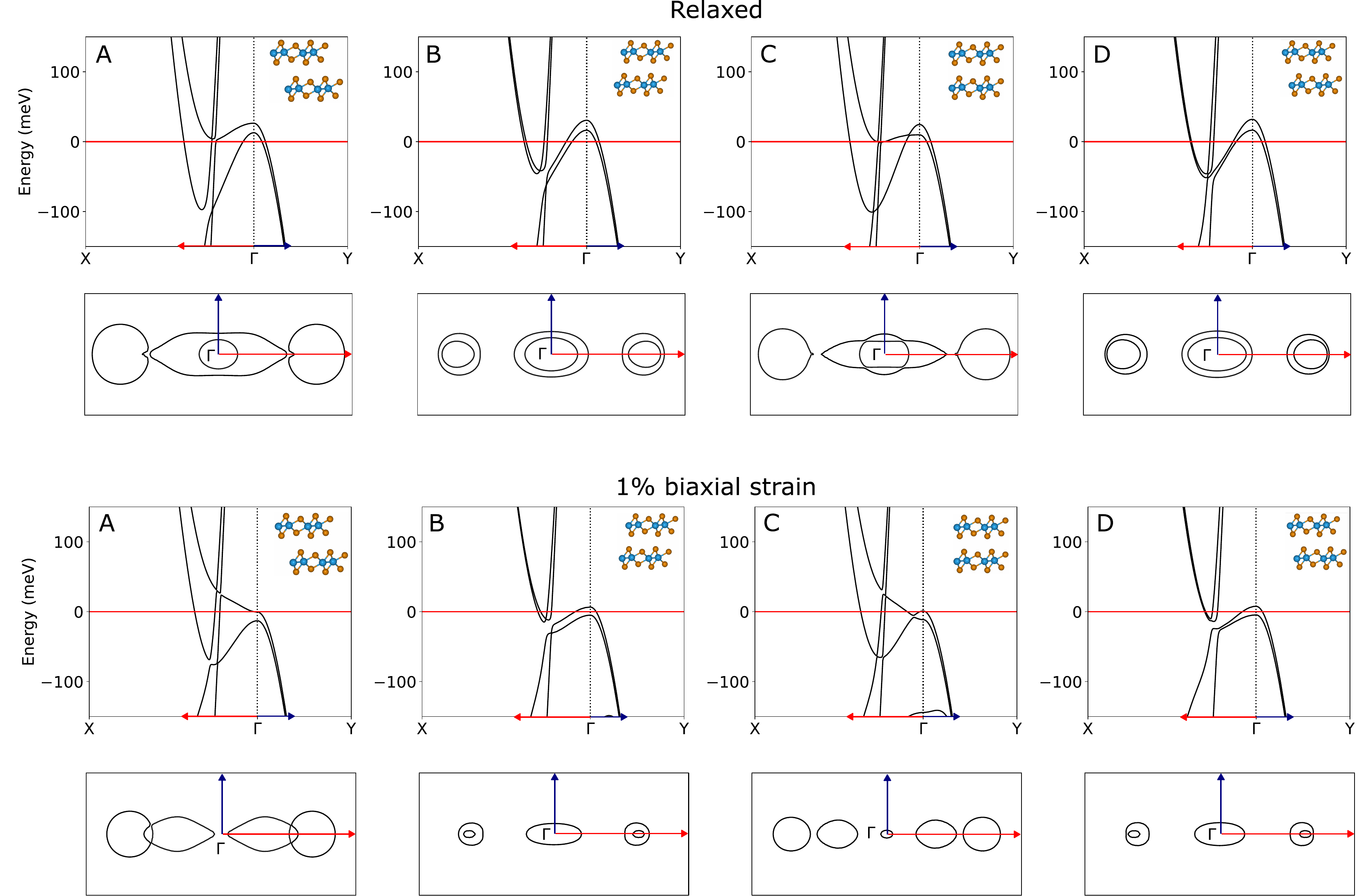}
    \caption{Band structures and Fermi surfaces, neglecting spin-orbit coupling, for commensurate aligned bilayers of 1T$'$ WTe$_2$, with relative in-plane shifts of the layers taken from the local atomic registries found at the four points indicated in Figs.~\ref{fig:structure} and \ref{fig:confining_potential}, and indicated schematically as insets. The interlayer distance is fixed as $d=7.75$~\AA. Top two rows are bands and Fermi surfaces calculated without strain, bottom rows are with 1\% in-plane biaxial strain applied.}
    \label{fig:aligned_bands}
\end{figure*}
Motivated by the appearance and strength of the confining potential on applying 1\% biaxial strain, as well as by the pattern in the stacking-dependence of interlayer hopping for aligned bilayers in Ref.~\onlinecite{Wu_WTe2}, we calculate using plane-wave DFT bands and Fermi surfaces for aligned bilayers with stackings corresponding to the local stackings at the points marked A-D in Figs.~ \ref{fig:structure}~and~\ref{fig:confining_potential} in the twisted bilayer.
These we plot in Fig.~\ref{fig:aligned_bands}.

While exactly at the top of the hole pocket at $\Gamma$ the interlayer hopping changes very little for the different stacking configurations, towards X within the electron pockets the changes are substantial.
In particular, the coupling changes sign on passing from configuration `A' to configuration `C', necessarily passing through zero in the region of configurations `B' and `D', where the layers become nearly electronically decoupled.
This gives rise to topoligical transitions in the Fermi surface character depending on the local stacking in different regions of the moiré supercell in the twisted bilayer.

Owing to the reduction of overlap between electron and hole pockets, here, as in the monolayer\cite{PhysRevLett.125.046801}, in the strained case these changes become more critical.
The $\Gamma$-point hole pocket dispersion becomes electron-like for the stacking (labelled `A') for which the chain of tungsten atoms in one layer lies above the midpoint between two chains in the other. 
These transitions and the flat bands associated with them provide an indication of the origin of the instability to the formation of a moiré potential wave in the twisted bilayer.

\section{Conclusions}
In conclusion, we have carried out a combination of linear-scaling and plane wave DFT calculations for a twisted bilayer of 1T$'$-WTe$_2$.
By comparing the electrostatic potential map of an aligned bilayer with those of isolated monolayers we find that the one-dimensional stripes present in the moiré pattern of the tungsten sublattices gives rise to an instability on application of a small strain to substantial confining potential wave with peak-to-trough magnitude $\sim$200~meV.
The effects of this confining striped moiré pattern are readily apparent in the unfolded spectral functions of the twisted bilayer.
Close examination of band structures from aligned bilayers with atomic registries corresponding to local stacking environments from different regions of the moiré superlattice shows notable changes to band structure and Fermi surface topologies depending on position in the moiré superlattice, particularly so for the slightly strained case. 
These computational results provide important theoretical underpinning for experimental results showing evidence of strong correlations in twisted 1T$'$-WTe$_2$ systems.
It should be noted that the imprecision of DFT both in the determination of structural parameters and in band alignment and bandwidth for a given structure means that the results which are closer to the behavior in a twisted bilayer studied in experiment may in fact be those which we have presented here as the `strained' case. Such twisted bilayers would therefore unstable to the formation of a confining potential wave.

While we were able to neglect atomic reconstruction as we considered a comparatively large twist angle, for very small twist angles we would expect highly non-uniform strain in the moiré supercell as the structure distorts to minimize its total energy\cite{Weston2020, PhysRevMaterials.8.034001}. Given the sensitivity of 1T$'$ WTe$_2$ to strain demonstrated here this would result in substantial changes to the electronic properties.
Materials isostructural to WTe$_2$ in which similar phenomena might be expected include 
the quasi-one-dimensional tellurides ${\mathrm{Ta}}_{1.2}{\mathrm{Os}}_{0.8}{\mathrm{Te}}_{4}$ \cite{JiaoPRB2022} and TaIrTe$_4$ \cite{XingTaIrTe4_2019,GuoTaIrTe4_2020}.

\section*{Acknowledgements}

SJM and NDMH acknowledge funding from EPSRC grant number EP/V000136/1. Computing facilities were provided by the Scientific Computing Research Technology Platform of the University of Warwick through the use of the High Performance Computing (HPC) cluster Avon, and the Sulis Tier 2 platforms at HPC Midlands+ funded by the Engineering and Physical Sciences Research Council (EPSRC), grant number EP/T022108/1. Computational support was also obtained from the UK national high performance computing service, ARCHER2, for which access was obtained via the UKCP consortium and funded by EPSRC grant ref EP/X035891/1.

\bibliography{references.bib}
\end{document}